\documentclass[aps,a4paper,twocolumn,showpacs]{revtex4}

\usepackage{amssymb}
\usepackage{amscd}
\usepackage{mathrsfs}
\usepackage{graphicx}

\newcommand{\be}   {\begin{equation}}
\newcommand{\ee}   {\end{equation}}
\newcommand{\bea}  {\begin{eqnarray}}
\newcommand{\eea}  {\end{eqnarray}}

\newcommand{\half} {\mbox {$\frac{1}{2}$}}

\begin{document}

\title{Phase Space Structure of Generalized Gaussian Cat States}

\author{Fernando Nicacio$^1$}
\email{nicacio@cbpf.br}

\author{Raphael N. P. Maia$^1$}

\author{Fabricio Toscano$^2$}
\email{toscano@if.ufrj.br}

\author{Ra\'ul O. Vallejos$^1$}
\email{vallejos@cbpf.br}
\homepage{http://www.cbpf.br/~vallejos}

\affiliation{ $^1$Centro Brasileiro de Pesquisas F\'{\i}sicas,
              Rua Dr.~Xavier Sigaud 150,
              22290-180 Rio de Janeiro, Brazil}

\affiliation{ $^2$Instituto de F\'{\i}sica,
              Universidade Federal do Rio de Janeiro,
              Cx.~P. 68528, 21941-972 Rio de Janeiro, Brazil}

\date{\today}

\begin{abstract}
We analyze generalized Gaussian cat states obtained by 
superposing arbitrary Gaussian states, e.g., a coherent 
state and a squeezed state.  
The Wigner functions of such states exhibit the typical 
pair of Gaussian hills plus an interference term which 
presents a novel structure, as compared with the standard 
superposition of coherent states (degenerate case). 
We prove that, in any dimensions, the structure of 
the interference term is characterized by a particular 
quadratic form; in one degree of freedom the phase is 
hyperbolic. 
This phase-space structure survives the action of a thermal 
reservoir. 
We also discuss certain superpositions of {\em mixed} Gaussian 
states generated by conditional Gaussian operations 
or Kerr-type dynamics on thermal states.
\end{abstract}


% 03.65.-w Quantum mechanics
% 03.65.Sq Semiclassical theories and applications in quantum mechanics
% 03.65.Yz Decoherence; open systems; quantum statistical methods
% 03.65.Ud Entanglement and quantum nonlocality (e.g. EPR paradox, 
%          Bell's inequalities, GHZ states, etc.) 

\maketitle

%%%%%%%%%%%%%%%%%%%%%%%
\section{Introduction}
\label{sec1}
%%%%%%%%%%%%%%%%%%%%%%%

The creation of quantum superpositions of living organisms,
what used to be merely a ridiculous possibility, may become 
feasible in a near future \cite{romero09}.
Of course, the center-of-mass states of a microbe, discussed 
by Romero-Isart {\em et al} \cite{romero09}, are still very 
distant from the fantastic dead/alive Schr\"odinger cat 
\cite{schrodinger35}.
In any case, for the time being, one will have to be satisfied 
with inorganic quantum superpositions which, 
however, may involve a large number of particles (``cat states").
At present they can be generated in the laboratory in a 
variety of systems, e.g., 
optical cavities             \cite{haroche}, 
superconducting devices      \cite{squid}, 
Bose-Einstein condensates    \cite{haroche}, 
free-propagating light beams \cite{ourjoumtsev07},
etc.
Not only are such states crucial for testing the 
nonlocality of quantum mechanics outside the microscopic
domain, but they also play an important role in some 
applications of quantum physics like 
information processing \cite{gilchrist04,infoproc}
and quantum metrology  \cite{gilchrist04,toscano06,dalvit06}.

The nonlocal quality of a cat state is most clearly 
displayed when it is looked through the Wigner 
representation.
For instance, in the case of a superposition of two 
coherent states, the Wigner function exhibits 
two Gaussian ``hills", 
corresponding to a classical superposition, 
and a nonclassical third Gaussian modulated by 
interference fringes halfway between the pair. 
This interference pattern is composed of alternating
positive and negative straight bands parallel to the 
line joining the two Gaussian peaks \cite{citation}.

The distributive property of the Wigner function allows 
one to construct the interference pattern of the
superposition of several (or many) coherent states 
\cite{zambrano09} from the corresponding Wigner function 
for each pair.
In fact, as the set of coherent states is complete,
any given pure state can be expressed 
as a coherent-state superposition: its Wigner 
function becomes a sum of cat-state Wigner functions.
This observation is sometimes enough to understand 
qualitative properties of general superpositions, 
like sensitivity to perturbations, or to the action
of the environment (decoherence) 
\cite{zurek01,jordan01}. 

In some proposed experiments the generated states 
consist of superpositions of coherent states lying on 
a manifold in phase space. 
This is the case, e.g., of the so-called arc states 
\cite{szabo94,tyc08}.
The Wigner function of such a state is characterized
by a banana-shaped positive region decorated by
interference fringes.
Reciprocally, a state whose Wigner function is 
localized in the vicinity of a phase-space manifold, 
can be fitted by a combination 
of coherent states concentrated on that manifold
\cite{kenfack03}.

Consider now the evolution of a superposition of 
coherent states under a nonlinear unitary dynamics.
In semiclassical regimes, for short times, an
individual coherent state moves along a classical
trajectory while suffering (approximately) a linear 
distortion. 
So, the wavepacket remains Gaussian but may 
become squeezed and rotated \cite{littlejohn86}.
In general, different coherent states of the 
superposition will be distorted in different ways.
This takes us to the following basic question: 
what happens to the Wigner function of a superposition 
of two coherent states when they are linearly, 
but independently, distorted?  
For instance, what is the interference structure 
of a superposition of a coherent state and a squeezed
state?

This paper is dedicated to the study of the coherence 
structure of generalized Gaussian cats in the Wigner
representation.
In the pure case, a Gaussian cat is, by definition, 
just a superposition of pure Gaussian states.
Even though there is no natural way of defining what a 
superposition of mixed states would be, there are
certain dynamics or protocols which produce what  
may be called mixed Gaussian cat states, given that
the Wigner function of such states are sums of 
Gaussians.
We analyze two classes of mixed cats: 
(i) pure cats that suffered the decohering action of a 
thermal environment, and
(ii) cats generated either by conditional Gaussian 
operations or by a Kerr-like dynamics on input thermal states. 
In all cases we focus on the geometric structure of
the interference fringes.

The paper is organized as follows.
We start by presenting our original dynamical 
motivation for studying generic superpositions of 
Gaussian states (Sect.~\ref{sec2}).
In Sect.~\ref{sec3} we summarize the geometrical 
formalism behind the Wigner representation which
will be instrumental for deriving most of the results
of this paper.
In Sect.~\ref{sec4} we analyze in detail the problem 
of a superposition of arbitrary pure Gaussian states.
We show that for one-degree of freedom (two dimensional
phase space) the interference pattern is in general
hyperbolic (instead of the linear pattern exhibited by
superposition of coherent states). 
For higher dimensions we derive a general normal form
which shows that, in a suitable canonical coordinate 
system, the fringe pattern is still hyperbolic in each
canonical plane.
Curiously, this pattern survives the action of a linear 
Markovian environment (Sect.~\ref{sec5}), although
its amplitude is progressively reduced over time.

Section~\ref{sec6} considers the result of using
standard schemes for generating pure cat states, but, 
instead of injecting a coherent-state at the input, one 
feeds the system with a thermal state. 
This also produces mixed cat states which, however,
are in general structurally different from the
decohered Gaussian cats of Sect.~\ref{sec4}.

Section~\ref{sec7} presents our concluding remarks.

%%%%%%%%%%%%%%%%%%%%%%%%%%%%%%%%%%%%%%%%%%%%%%%%%%%
\section{A motivation from semiclassical dynamics}
\label{sec2}
%%%%%%%%%%%%%%%%%%%%%%%%%%%%%%%%%%%%%%%%%%%%%%%%%%%

A typical method for the approximate semiclassical propagation
of quantum states consists of decomposing the initial 
state (possibly delocalized) into a suitable superposition 
of coherent states which are themselves propagated 
approximately by taking advantage of their quasiclassical 
nature \cite{shalashilin08}. 

In the crudest scheme, each coherent state moves along a 
classical trajectory while preserving its shape
(frozen Gaussian approximation). 
In the next level of improvement, an initial coherent state 
stays Gaussian during its evolution but may become squeezed 
and rotated 
(thawed Gaussian approximation) \cite{tannor}.
Both schemes correspond to approximating the exact Hamiltonian
by its Taylor expansion around the instantaneous center of
the wavepacket $(q(t),p(t))=x(t)$:
\begin{eqnarray}
H( \hat x,t ) &=& H(x(t))
+ (\hat x -x(t)) \cdot \frac{\partial H}
            {\partial \hat x}\bigg|_{x(t)} \nonumber \\
&+& (\hat x -x(t))
\cdot \frac{\partial^2 H}
           {\partial \hat x^2}\bigg|_{x(t)} \!\!
           (\hat x -x(t))  + ...                             
\end{eqnarray}
Here $\hat x$ represents the pair of canonical operators 
$(\hat q,\hat p)$.
The frozen/thawed schemes are obtained by truncating the Taylor
expansion to first/second order in $\hat x$
\cite{littlejohn86,tannor}. 
In both cases the dynamics generated by $H(\hat x,t)$ is
linear.

\begin{figure}[htp]
\hspace{0.0cm}
\includegraphics[angle=-90.0, width=8cm]{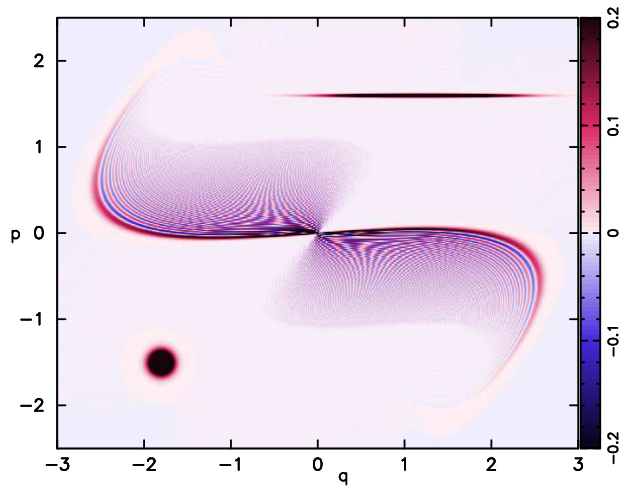}
\caption{%
(color online) 
Nonlinear dynamics in the kicked harmonic oscillator.
Shown are Wigner functions of 
(a) the evolved state at $t=2^-$;
(b) the initial squeezed state ($t=0^-$), which is centered 
at the origin but has ben displaced for clarity;
(c) a coherent state, displayed for reference.
Parameters are 
$K = 2$, 
$\tau = \pi/3$, 
$\hbar= 0.0128 $.}
\label{fig1}
\end{figure}

Let us illustrate the semiclassical schemes above with a 
numerical example.
Consider the propagation of a squeezed state in the
Kicked Harmonic Oscillator (KHO).
In suitable units the KHO Hamiltonian reads 
\cite{KHO}:
\begin{equation}
H(\hat q, \hat p,t) =
\frac{1}{2} \left( \hat p^2 + \hat q^2 \right)
+ K \cos( \hat q )
\sum_{n = -\infty}^{\infty} \delta( t - n \tau ).   
\label{HKHO}                     
\end{equation}
We choose an initial state $|\psi_0 \rangle$ which is 
centered at the phase-space origin and squeezed along the 
$q$-axis. 
By appropriate choice of the parameters $\tau$ and 
$K$, strong nonlinear effects can be clearly seen after 
two pulses (kicks). 
See Fig.~\ref{fig1}.
%
%
%%%%%%%%%%%%%%%%%%%%%%%%%%%%%%%%%%%%%%%%%%%%%%%%%%%%%%%%%%%%%
%%%%%%%%%%%%%%%%%%%%%%%%%%%%%%%%%%%%%%%%%%%%%%%%%%%%%%%%%%%%%
%
%%%%%%%%%%%%%%%%%%%%%%%%%%%%%%%%%%%%%%%%%%%%%%%%%%%%%%%%%%%%%
%%%%%%%%%%%%%%%%%%%%%%%%%%%%%%%%%%%%%%%%%%%%%%%%%%%%%%%%%%%%%
%
%

\begin{figure}[htp]
\hspace{0.0cm}
\includegraphics[angle=0.0, width=7cm]{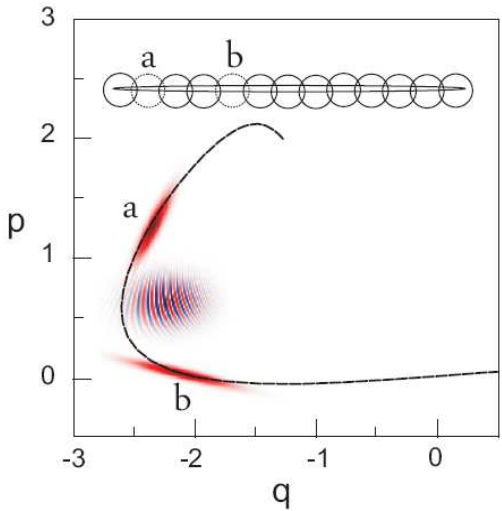}
\caption{%
(color online) 
The initial squeezed state of Fig.~\ref{fig1} is 
decomposed into coherent states lying on the $q$-axis
(top, schematic).
Then each coherent state is evolved according to the 
local linear dynamics, so that it preserves its Gaussian 
shape but may rotate and get squeezed. 
Two examples are shown (a,b), together with their joint 
cat-state Wigner function.
At $t=2^-$ the $q$-axis has evolved into the curved 
manifold.
Same parameters as in Fig.~\ref{fig1}.}
\label{fig2}
\end{figure}

\begin{figure}[htp]
\hspace{0.0cm}
\includegraphics[angle=0.0, width=9cm]{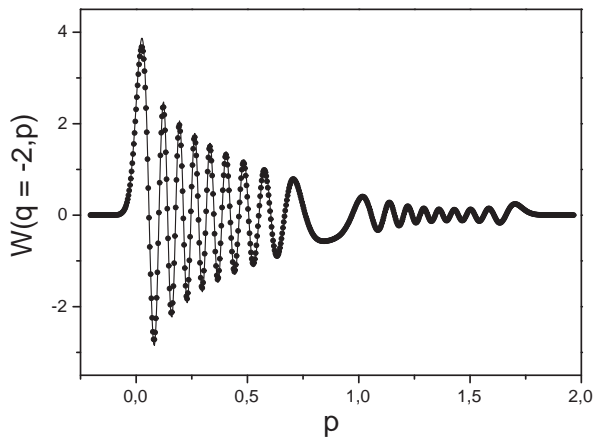}
\caption{%
Exact (thin line) vs. approximate (circles) Wigner functions
at time $t=2^-$.
We exhibit the section $q=-2$ of the Wigner function, 
i.e., $W(q=-2,p)$ vs. $p$.
Same parameters as in previous figures.}
\label{fig3}
\end{figure}

Our approximate description of the final state starts by 
decomposing the initial squeezed state $|\psi_0 \rangle$ 
in terms of a one-parameter family of coherent states 
$|\phi(q')\rangle$ with centers lying on the line $p=0$:
\begin{equation}
 |\psi_0 \rangle = 
 \int_{-\infty}^{+\infty} dq' \, C(q') \, |\phi(q')\rangle \,,
\end{equation}
where $\langle q |\psi_0 \rangle$, $\langle q |\phi(q')\rangle$
and $C(q')$ are real Gaussians \cite{szabo96} 
(in numerical calculations the integral is substituted by
a finite sum).
Then we propagate each $|\phi(q')\rangle$ using the instantaneous
local quadratic Hamiltonian, i.e., in Eq.~(\ref{HKHO}) we make
the approximation
\begin{equation}
\cos( \hat q) \approx a(t) + 
                      b(t) \, \hat{q} + 
                      c(t) \, \hat{q}^2  \, ,
\end{equation}
with $a(t)= \cos(q(t))$, etc. 
In this way the final state becomes a superposition of Gaussian 
states, their centers threaded by a curved manifold in phase 
space (see Fig.~\ref{fig2}).
% 
%
%%%%%%%%%%%%%%%%%%%%%%%%%%%%%%%%%%%%%%%%%%%%%%%%%%%%%%%%%%%%%
%%%%%%%%%%%%%%%%%%%%%%%%%%%%%%%%%%%%%%%%%%%%%%%%%%%%%%%%%%%%%
%
%%%%%%%%%%%%%%%%%%%%%%%%%%%%%%%%%%%%%%%%%%%%%%%%%%%%%%%%%%%%%
%%%%%%%%%%%%%%%%%%%%%%%%%%%%%%%%%%%%%%%%%%%%%%%%%%%%%%%%%%%%%
%
%
Figure~\ref{fig3} presents a numerical comparison 
between approximate and exact Wigner functions.
The excellent agreement indicates that the global 
nonlinear dynamics of the initial state can be understood,
both qualitatively and quantitatively,
as the collective effect of a swarm of localized states 
evolving linearly along classical trajectories.
%
%
%%%%%%%%%%%%%%%%%%%%%%%%%%%%%%%%%%%%%%%%%%%%%%%%%%%%%%%%%%%%%
%%%%%%%%%%%%%%%%%%%%%%%%%%%%%%%%%%%%%%%%%%%%%%%%%%%%%%%%%%%%%
%
%%%%%%%%%%%%%%%%%%%%%%%%%%%%%%%%%%%%%%%%%%%%%%%%%%%%%%%%%%%%%
%%%%%%%%%%%%%%%%%%%%%%%%%%%%%%%%%%%%%%%%%%%%%%%%%%%%%%%%%%%%%
%
%

The interference pattern of the Wigner function of 
Fig.~\ref{fig1} can now be viewed as arising from
the superposition of the elementary patterns for
each pair of final Gaussian states.
In principle each wavepacket suffers a different 
distortion, so, the cat states obtained by superposing
pairs of wavepackets are in general nonstandard in the
sense that their interference fringes are not parallel.
Inspection of Fig.~(\ref{fig2}) suggests that 
in the general case the interference fringes may be 
hyperbolic.

%The scheme above is not limited to initial squeezed states.
%If dynamics is nonlinear, even initial coherent states can 
%be propagated provided one waits for a while ...

In Sect.~\ref{sec4} below we formulate and prove a precise 
statement about the hyperbolicity of the interference pattern 
of the superposition of two arbitrary pure Gaussian states. 
But first we need to review some mathematical results and
notations which are extremely useful when
dealing with Wigner functions and their related characteristic 
functions.

%%%%%%%%%%%%%%%%%%%%%%%%%%%%%%%%%%%
\section{Mathematical Formalism}
\label{sec3}
%%%%%%%%%%%%%%%%%%%%%%%%%%%%%%%%%%%

Most of the formulae of this section are widespread 
in the literature, sometimes under different notations
\cite{littlejohn86,ozorio98,degosson,mehlig01}.
For the sake of self-completeness we present them 
compactly using a unified notation.
The idea of the present formalism is to exploit the geometry
of translations and reflections behind the Wigner
representation \cite{ozorio98}.
In this way, many of the cumbersome integrals
usually needed in Wigner calculus
are reduced to simple algebra.

%%%%%%%%%%%%%%%%%%%%%%%%%%%%%%%%%%%%%%%%%
\subsection{Translations and Reflections}
%%%%%%%%%%%%%%%%%%%%%%%%%%%%%%%%%%%%%%%%%

Given a quantum state described by the density operator 
$\hat \rho$, its Wigner function is defined by 
\begin{equation}
W(x) =
      \frac{1}{(2 \pi \hbar)^{n}} 
      \int^{+\infty}_{-\infty}  
      dq' \, 
      \langle q + \textstyle \frac{1}{2} q'|
      \, \hat{\rho} \,
      | q - \textstyle \frac{1}{2} q' \rangle \,
      e^{- i p \cdot q'/\hbar}.                  
\label{Wigner}
\end{equation}
We assume that the system possesses $n$ degrees of freedom.
Canonical coordinates will be represented collectively by 
the column vectors
$q=(q_1,\ldots,q_n)^\top$ 
and 
$p=(p_1,\ldots,p_n)^\top$
(of course, $\top$ means ``transposed").
A point in phase space is characterized by a $2n$-dimensional
column vector 
$x=(q_1,\ldots,p_n)^\top$.
A similar notation is used for operators
$\hat q$, $\hat p$, and $\hat x$, e.g., 
$\hat q=(\hat q_1, \ldots , \hat q_n)^\top$.
(As an alternative to the mechanical point of view, 
one may adopt an optical perspective: 
If $\hat \rho$ represents the state of 
an $n$-mode quantum radiation field, then $q$ and $p$ are 
conjugate quadrature-vectors of the field.)

Closely related to the Wigner function, the 
(symmetric ordered) characteristic function $\chi(\xi)$ 
is defined as the expectation value of the 
phase-space translation operator:
\begin{equation}
\chi(\xi) = \frac{1}{(2 \pi \hbar)^{n}} \,
            \mbox{tr} 
            \left( {\hat \rho} \, \hat{T}_\xi^\dagger \right) \,,
\label{chiofksi}
\end{equation}
with $\xi=(\xi_q^\top,\xi_p^\top)^\top$, and
\begin{equation}
 \hat{T}_\xi = e^{i \, \xi \wedge \hat{x}/\hbar} \,.
\label{txi}
\end{equation}
In the last equation we have introduced the wedge (symplectic)
product:  
\begin{equation}
\xi \wedge \hat{x} = \xi_p \cdot \hat{q} - \xi_q \cdot \hat{p}  \,,
\end{equation}
which can be turned into an ordinary scalar product 
(or matrix product) with the help of the symplectic 
matrix $\mathsf J$,
\begin{equation}
\xi \wedge \hat{x} = 
(\mathsf J \, \xi) \cdot \hat x =
\xi^\top  \mathsf J^\top \hat x  \,,
\end{equation}
where $\mathsf J$ is given by  
\begin{equation}
\mathsf J = \left(
             \begin{array}{cc}
                     \mathsf 0_n & \mathsf I_n \\
                    -\mathsf I_n & \mathsf 0_n
              \end{array}
    \right) \,.
\end{equation}
Here 
$\mathsf I_n$ and $\mathsf 0_n$ are the $n$-dimensional 
identity and null matrix respectively.
A symplectic matrix $\mathsf S$ is defined by the property 
of preserving the wedge product, what amounts to
$ \mathsf S \mathsf J \mathsf S^\top = \mathsf J$.

The Weyl-Heisenberg translation operator $\hat{T}_\xi$ is 
equivalent to Glauber's optical displacement 
operator $\hat{D} (\alpha)$,
which is expressed in terms of a complex parameter $\alpha$ 
and the annihilation/creation operators 
$\hat{a}/\hat{a}^\dagger$,
\be
\hat{D} (\alpha)=
\exp(\alpha \, \hat{a}^\dagger - \alpha^\ast \hat{a}) \, .
\ee
The correspondence is established by making the 
identifications 
$\alpha=(\xi_q+i\xi_p)/\sqrt{2 \hbar}\,$ 
and 
$\hat{a}=(\hat{q}+i\hat{p})/\sqrt{2 \hbar}$
\cite{hillery84,barnett}.

Though not so widely known, the Wigner function can also be 
written as an expectation value:
\begin{equation}
W(x) = \frac{1}{( \pi \hbar)^{n}} \,
            \mbox{tr} 
            \left( {\hat \rho} \, \hat{R}_x \right) \,.
\label{wofx}
\end{equation}
The (Grossman-Royer \cite{degosson}) operators $\hat{R}_x$ 
are both unitary and hermitian. 
In particular, $\hat{R}_{x=0}$ coincides with the usual parity 
operator, i.e.,
$\hat{R}_0 |q\rangle=|-q \rangle$.
Given that the parity operator corresponds to a geometrical 
reflection through the phase-space origin, and taking into
account the property
\be
\hat T_x \hat R_0 \hat T_x^\dagger = \hat R_x \, ,
\ee
the operator $\hat{R}_x$ is interpreted naturally as 
the quantum version of the phase-space reflection through 
the point $x$ \cite{ozorio98}.

Both sets of reflections and translations,
$\{\hat R_x\}$ and $\{\hat T_\xi\}$, 
constitute orthogonal bases in operator space
(with respect to the Hilbert-Schmidt
product).
Thus any operator $\hat A$ can be written as a linear 
combination of reflections or translations \cite{ozorio98}:
\bea
\label{intAR}
\hat{A} & = & \frac{1}{ ( 2 \pi\hbar )^n }
              \int^{+\infty}_{-\infty} 
              dx \, A(x) \, \hat R_x \, , \\
        & = & \frac{1}{ ( 2 \pi\hbar )^n }
              \int^{+\infty}_{-\infty} 
              d\xi \, \tilde{A}(\xi) \, \hat T_\xi \,,    
\eea
where the reflection and translation symbols, 
$A(x)$ and $\tilde{A}(\xi)$ respectively, are
given by:
\bea
\label{Aofx}
A (x)           & = & 2^n \, {\rm tr}  
            \left( \hat A \, \hat R_x   \right) \, , \\
\tilde{A} (\xi) & = &       {\rm tr} 
            \left( \hat A \, \hat T_\xi^\dagger \right) \, . 
\label{Aint}      
\eea
The symbols $\tilde A(\xi)$ and $A(x)$ are related
via a (symplectic) Fourier transformation
\begin{equation}
A(x) = \frac{1}{(2\pi\hbar)^n }
       \int d \xi \, \tilde A(\xi)\,
        e^{ i \, \xi \wedge x/\hbar } \, .                      
\end{equation} \\
The relation between the Wigner and characteristic 
functions is a particular case ($\hat A= \hat \rho$) 
of the general expression above:
\begin{equation}
W(x) = \frac{1}{(2\pi\hbar)^n }
       \int d \xi \,  \chi(\xi)\,
        e^{ i \, \xi \wedge x/\hbar } \, .                      
\end{equation} \\
%

%%%%%%%%%%%%%%%%%%%%%%%%%%%%%%%%%%%%%%%%%
\subsection{Metaplectic transformations}
%%%%%%%%%%%%%%%%%%%%%%%%%%%%%%%%%%%%%%%%%

The quantum analog of a linear canonical (symplectic)
transformation $\mathsf{S}$ is a metaplectic unitary 
operator 
$\hat M_\mathsf{S}$ \cite{littlejohn86}.
We have already seen an example: 
the parity operator $R_0$ is the quantum version
of the classical reflection 
$\mathsf S = -\mathsf I_{2n}$.
A metaplectic operator can be thought as the quantum propagator
associated with a Hamiltonian that is purely quadratic in the 
canonical operators $\hat q$ and $\hat p$
\cite{littlejohn86, ozorio98}.
The metaplectic operators together with the translation 
operators (associated with Hamiltonians that are purely linear 
in $\hat q$ and $\hat p$),
constitute the set of unitary Gaussian operations, i.e.,
operations that leave the set of Gaussian states invariant
\cite{braunstein05}.

Metaplectic operators respect the classical group composition 
law, i.e., if $\mathsf{S}$ and $\mathsf{S'}$ are symplectic
transformations, then  
\begin{equation}
\hat M_{ \mathsf{S S'} } = \pm \,
\hat M_\mathsf{S} \hat M_\mathsf{S'} \, ,
\label{MSS}
\end{equation}
where the $\pm$ sign (unessential for our purposes) 
depends on both 
$\mathsf{ S , S'}$ \cite{degosson, littlejohn86}.

Metaplectics, translations and reflections interact 
according to the following formulas
\cite{littlejohn86, ozorio98}:
\begin{eqnarray}
\label{TRT}
\hat T_\xi \hat R_x \hat T_\xi^\dagger &=& 
\hat R_{ x + \xi }             \, ,      \\
\label{MTM}
\hat M_\mathsf{S} \hat T_\xi \hat M_\mathsf{S}^\dagger &=&
\hat T_{ \mathsf S \xi }       \, ,      \\
\label{MRM}
\hat M_\mathsf{S} \hat R_x \hat M_\mathsf{S}^\dagger &=&
\hat R_{ \mathsf S x }         \, .    
\end{eqnarray} 
These expressions entail the covariance of the Wigner
function with respect to both translations and metaplectic
transformations. 
When a state is translated, 
$\hat \rho^\prime =\hat T_\xi \hat \rho T_\xi^\dagger$,
one deduces from Eq.~(\ref{TRT}) that its Wigner
function is also translated:
\be
W_{\rho^\prime}(x)= W_\rho (x-\xi) \, ,
\label{TWT}
\ee
using that 
$\hat T_\xi^\dagger=\hat T_{-\xi}$,
and the definition (\ref{wofx}).
Analogously, for a metaplectically deformed state,
$\hat \rho^\prime =
 \hat M_\mathsf{S} \hat \rho \hat M_\mathsf{S}^\dagger$,
one easily proves the metaplectic covariance of the Wigner 
and the characteristic functions
\bea
\label{MWM}
   W_{\rho^\prime}( x ) & = &    W_\rho (\mathsf{S}^{-1} x ) \, , \\
\label{MXM}
\chi_{\rho^\prime}(\xi) & = & \chi_\rho (\mathsf{S}^{-1}\xi) 
\eea
[using 
$\hat M_\mathsf{S}^\dagger=\hat M_{\mathsf{S}^{-1}}$,
and Eqs.~(\ref{chiofksi},\ref{wofx},\ref{MTM},\ref{MRM})].

The reflection symbol 
$M_\mathsf{S}(x)$
% \propto {\rm tr} (\hat M_\mathsf{S} \hat R_x)$
%
is an essential ingredient in our calculations of 
Sec.~\ref{sec4}.
It is obtained from (\ref{Aofx}) and given by 
\cite{ozorio98,mehlig01,degosson}
\begin{equation}
M_\mathsf{S} (x) =
    \frac{2^n \, i^\nu}
         {\sqrt{\left|\det( \mathsf S + \mathsf{I}_{2n} )
               \right|}}  
     \exp  \left( \frac{i}{\hbar} \,
                  x \cdot \mathsf C_\mathsf{S} \, x 
           \right) \, ,     
\label{Mofx}                   
\end{equation}
where the symmetric matrix $\mathsf C_\mathsf{S}$
is the Cayley transform of the symplectic
matrix $\mathsf S$ \cite{ozorio98,degosson}:
\begin{equation}
\mathsf C_\mathsf{S} = \mathsf J
\left( \mathsf{S} - \mathsf{I}_{2n} \right)
\left( \mathsf{S} + \mathsf{I}_{2n} \right)^{-1}   \, .             
\end{equation}
An explicit expression for the integer $\nu$ can be 
found in Ref.~\cite{degosson}.
Like the $\pm$ sign in Eq.~(\ref{MSS}), 
$\nu$ is a delicate object, and not really necessary 
for our dicussions. 
For these reasons we shall omit all details
about this phase.

Even when $\mathsf{S} + \mathsf{I}_{2n}$ is singular,
Eq.~(\ref{Mofx}) is still meaningful if we interpret
that formula as a limit (in which one of the 
eigenvalues of $\mathsf{S}$ tends to $-1$) 
\cite{littlejohn86}.

In principle we could have worked in the $q$-representation. 
However, the expression for $M_\mathsf{S} (q,q^\prime)$
requires the explicit splitting of $\mathsf{S}$ into four blocks
\cite{littlejohn86}.
This would lead to extremely cumbersome expressions when 
considering {\em pairs} of metaplectic transformations, like
in Sec.~\ref{sec4} below.

%%%%%%%%%%%%%%%%%%%%%%%%%%%%%
\subsection{Gaussian States}
%%%%%%%%%%%%%%%%%%%%%%%%%%%%%

According to the standard definition, a state is
said Gaussian if its Wigner function is a Gaussian.
Any pure Gaussian state can be obtained from a
fiducial Gaussian state by the combined action of a 
translation and a metaplectic transformation \cite{degosson}.

For simplicity, we take as fiducial state the ground 
state of the $n$-dimensional isotropic harmonic oscillator.
It will be denoted $|0\rangle$.
In a system of units where frequency and mass
are unity, $\omega = m = 1$, the position wavefunction
reads:
\begin{equation}
\langle q | 0 \rangle = \frac{1}{ (\pi\hbar)^{n/4} } \,
                        e^{ -q^2/2 \hbar} \,  .           
\label{fiducial}
\end{equation}
A general pure Gaussian state can be obtained 
from $|0\rangle$ by 
the successive application of a metaplectic operator 
and a translation:
\be
 |\mathsf S, \zeta \rangle =
      \hat T_\zeta \hat M_\mathsf{S} | 0\rangle \, .   
\label{general}                   
\ee
For the sake of compactness of forthcoming formulae,
let us define a normalized Gaussian function:
\be
\mathcal G 
\left( x; \mathsf M, \zeta \right) =
\frac{\sqrt{\det \mathsf M}}{(\pi\hbar)^n}
     \exp \left[ - \left( x - \zeta \right)  \cdot
                           \mathsf M
                   \left( x - \zeta \right)/\hbar
         \right] \, ,
\ee
where $x$ and $\zeta$ are $2n$-dimensional real
vectors, and $\mathsf M$ is a symmetric 
$2n \! \times \! 2n$
complex matrix with positive real part, 
$\rm{Re} \mathsf M > 0$. 
The branch of square root is chosen in such a way that 
it reduces continuously to the positive root in the
case of real $\mathsf M$ \cite{folland}.
Note that if $\mathsf M$ is symplectic 
(a case we shall meet several times), then 
$\det \mathsf M=1$ \cite{mackey03}. 

When $\mathcal G \left( x; \mathsf M, \zeta \right)$
represents a probability distribution $\mathsf M^{-1}$ 
is the {\em covariance} matrix. 
Sometimes we shall abuse of the language and 
use such denomination for arbitrary Gaussians; 
in that cases ``covariance matrix" is to be understood 
as a shorthand for ``the inverse of the matrix of the
quadratic form ...".

The Wigner function of the fiducial state (\ref{fiducial}) 
is readily calculated:
\be
 W_0(x) = \mathcal G  \left(x; \mathsf I, 0 \right) \, .
\ee
The Wigner function of the general Gaussian state 
$|\mathsf S, \zeta \rangle $ is obtained from the fiducial 
one by using the covariance properties (\ref{TWT},\ref{MWM}). 
The result is
\be
W_{|\mathsf S, \zeta \rangle }(x) = \mathcal G
  \left[ x;  (\mathsf S \mathsf S^{\top})^{-1} , \zeta 
 \right] \, . 
\label{W0}                                                  
\ee
In this case, the covariance matrix 
$\mathsf S \mathsf S^{\top}$
is symplectic.
It is worth showing also the corresponding characteristic 
function:
\be
\chi_{|\mathsf S, \zeta \rangle }(\xi) = 2^{-n} \,
\mathcal G 
\left[ \xi/2 ; (\mathsf S \mathsf S^\top)^{-1} , 0 
\right] \, e^{i\zeta \wedge \xi/\hbar}  \, . 
\label{chi0}      
\ee
Note that, for pure Gaussian states, 
$W(x)$ coincides with $\chi(2x)$, 
except for a factor and a shift of origin.

%%%%%%%%%%%%%%%%%%%%%%%%%%%%%%%%%%%%
\section{Gaussian Cat States}
\label{sec4}
%%%%%%%%%%%%%%%%%%%%%%%%%%%%%%%%%%%%
%
We shall call ``Gaussian cat state" any
superposition of two pure Gaussian states, i.e.,
\begin{equation}
| \textstyle \Psi \rangle =
              a \, | \mathsf U, u \rangle  +
              b \, | \mathsf V, v \rangle     \, ,    
\label{Psi}         
\end{equation}
where $a$ and $b$ are complex amplitudes, arbitrary
to the extent that the state remains normalized.
Its Wigner function reads
\be
W_\Psi(x) =   |a|^2 \, W_{| \mathsf U, u \rangle}(x) +  
              |b|^2 \, W_{| \mathsf V, v \rangle}(x) +  
              |ab| \, \mathcal{I}(x) \, .
\label{cat}
\ee
The Wigner function of the superposition is the
sum of the individual Wigner functions 
(two Gaussian ``hills" centered at $u$ and $v$) 
plus an interference term, given by
\be
\mathcal{I}(x)= 
\frac{2}{(\pi \hbar)^n} \,
{\rm Re} \left[
e^{i\varphi}
\langle {\mathsf U}, u | \, \hat R_x \, |\mathsf V, v \rangle 
\right] \, ,
\end{equation}
with $\varphi=\arg(a^\ast b)$. 
A typical example of the superposition (\ref{cat}) was
depicted in Fig.~\ref{fig2}.

The purpose of this section is to explore the structure of
the interference term.
We start by writing the matrix element in the last
equation as a vacuum expectation value:
\be
\langle \mathsf U, u | \hat R_x |\mathsf V, v \rangle  =
\langle 0 |
\hat M_\mathsf{U}^\dagger \hat T_u^\dagger
\hat R_x
\hat T_v \hat M_\mathsf{V}
| 0 \rangle   \,.                                                 
\ee
First we use the fact that the composition of a 
translation and a reflection is also a reflection
\cite{ozorio98}, 
\bea
\label{RTR}
\hat R_x \hat T_\xi & = &  
             e^{i \xi \wedge x / \hbar} 
             \, \hat R_{x-\xi/2} \, ,         \\
\label{TRR}
\hat T_\xi \hat R_x & = &  
             e^{i \xi \wedge x / \hbar} 
             \, \hat R_{x+\xi/2} \, ,
\eea
in order to transform 
$\hat T_u^\dagger \hat R_x \hat T_v$
into a single reflection times a phase.
After invoking Eqs.~(\ref{MSS},\ref{MRM}) we arrive at
\be
\langle \mathsf U, u | \hat R_x |\mathsf V, v \rangle  =
e^{ i     x  \wedge \zeta /   \hbar
+   i  \zeta \wedge \eta  / 2 \hbar } \,
\langle 0 |
\hat R_{\mathsf U^{-1 } \left( x - \eta \right) }
\hat M_{\mathsf U^{-1 } \mathsf V}
| 0 \rangle \, ,                                                   
\ee
with
\bea
\zeta & = & u - v        \, , \\
\eta  & = & (u + v)/2    \, .
\eea

We sketch the next steps skipping the details:
(i) Expand $\hat M_{\mathsf U^{-1 } \mathsf V}$ 
into reflection operators
with the help of Eqs.~(\ref{intAR},\ref{Mofx}).
(ii) Turn the resulting product of reflections into a 
translation using the composition formula \cite{ozorio98}
\be
\label{RRT}
\hat R_x \hat R_y  =  
                        e^{2 i y \wedge x / \hbar} 
             \, \hat T_{2(x-y)} \, .         
\ee
(iii) At this point one must calculate the average
$ \langle 0 | \hat T_\xi | 0 \rangle$, for a certain 
$\xi$. 
But this is essentially the characteristic function of 
the vacuum, which can obtained from Eq.~(\ref{chi0}) by
setting $\mathsf S = \mathsf I$ and $\zeta=0$.
(iv) Calculate the remaining Gaussian integral.
The final result is
\be
\langle 0 |
\hat R_{\mathsf U^{-1} \left( x - \eta \right) }
\hat M_{\mathsf{U}^{-1}\mathsf V}
| 0 \rangle
=
(\pi\hbar)^n \, K \,
\mathcal{G} \left( x ; \mathsf G , \eta \right) \, ,
\ee
with
\bea
K & = &  \frac{ 2^n \, i^\mu }{
                 \sqrt{ 
          \det \left[
       \left( \mathsf U + \mathsf V \right)  + 
    i  \left( \mathsf U - \mathsf V \right) \mathsf J
              \right] }        }  \, , \\
\mathsf G & = &
      \left(
            \mathsf{ U U^{\top}}  + 
            \mathsf{ V V^{\top}}
      \right)^{-1}  
        \left[ 2-i
      \left( \mathsf{ U U^{\top}} - 
             \mathsf{ V V^{\top}}
      \right) \mathsf J   
       \right] \, ,                                           
\eea
and $\mu$ an integer.

Summing up, the interference term may be written as
\be
\mathcal{I}(x) = 
2 \, | K| 
\, {\rm Re} \left[
e^{i x \wedge \zeta/\hbar + i \phi } \,
\mathcal{G} \left( x ; \mathsf G , \eta \right) \right] \, .  
\label{Iofx2}                    
\ee
Here $\phi$ stands for a phase that does not depend 
on $x$. 
Its precise value is not relevant for the forthcoming
analysis.

It can be checked that $\mathsf G$ is not only symmetric
but is also complex symplectic, i.e., 
$\mathsf G \mathsf J \mathsf G^{\top} = \mathsf J$.
Then, one has $\det \mathsf G = 1$ \cite{mackey03}.

%%%%%%%%%%%%%%%%%%%%%%%%%%%%%%%%%%%%%%%%%%%%%
\subsection{Geometry of the Wigner Function}
%%%%%%%%%%%%%%%%%%%%%%%%%%%%%%%%%%%%%%%%%%%%%

The interference pattern (\ref{Iofx2}) can be factored 
into a positive Gaussian envelope times an oscillatory 
function.
The envelope is given by:
\be
\mathcal{I}_{\rm env}(x)=
\frac{ 2 \, | K|}{(\pi \hbar)^n} \,
\exp 
 \left[ -( x -\eta) \cdot {\rm Re} \, \mathsf G \, (x-\eta)/\hbar
\right] \, .  
\label{Ienv}                    
\ee
This Gaussian is centered at $x=\eta$, the midpoint
between the centers of the individual Wigner distributions
$W_{|\mathsf U, u \rangle}(x)$ 
and 
$W_{|\mathsf V, v \rangle}(x)$.
Its covariance matrix, $(\rm{Re} \mathsf G)^{-1}$, is an 
average of the individual covariance matrices:
\be
(\rm{Re} \mathsf G)^{-1} =   
          \left(\mathsf{ U U^{\top}}  + 
                \mathsf{ V V^{\top}}
          \right)/2  \, .                   
\ee
If $\mathsf U = \mathsf V$ 
(equally distorted wavepackets), then the shape 
(covariance matrix) of the Gaussian envelope is equal 
to the individual Wigner functions. 
The height of the envelope depends on the amplitudes
$a$ and $b$ of the Gaussian states [see Eq.~(\ref{Psi})], 
e.g., for equal amplitude superpositions, $|a|=|b|$, 
the Gaussian envelope (\ref{Ienv}) is twice as high as 
the individual Wigner functions. 

\begin{figure*}[htp]
\hspace{0.0cm}
\includegraphics[angle=-90.0, width=15cm]{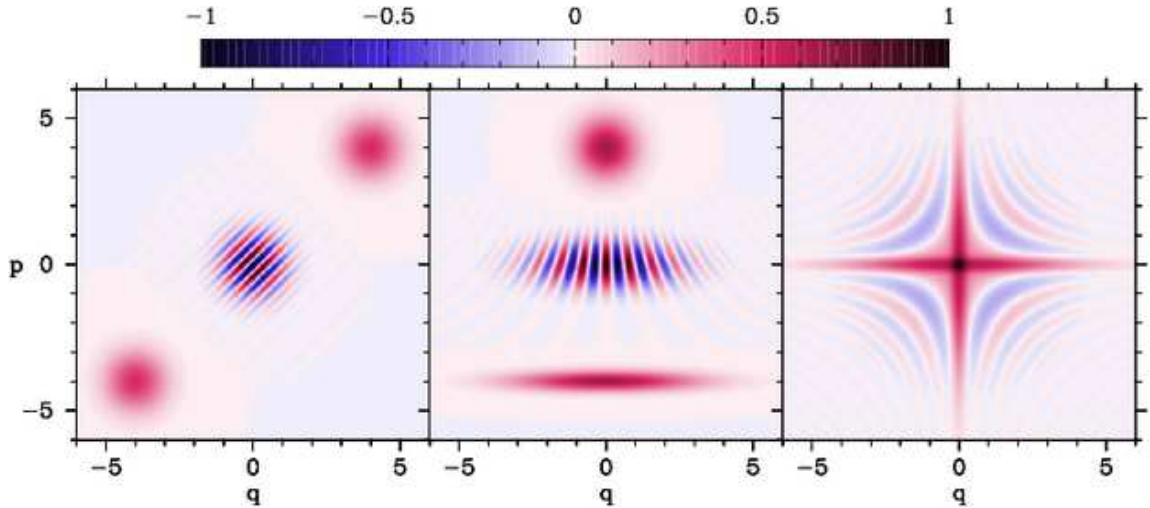}
\caption{(color online)
A gallery of Gaussian-cat-state Wigner functions.
We show equal-weight superpositions, i.e., 
Eq.~(\ref{Psi}) with $|a|=|b|$.
>From left to right:
(a) 
Standard superposition of coherent states displaying
a linear interference pattern. 
(b) Superposition of a squeezed state and a coherent
state.
(c) Superposition of two squeezed states both centered
at the origin and with orthogonal squeezing directions.}
\label{fig4}
\end{figure*}

The oscillatory part reads
\be
\label{Iosc}
{\mathcal I}_{\rm osc}(x)=
\cos  \left[ 
\phi +  
x \wedge \zeta/\hbar +
(x-\eta) \cdot {\rm Im} \, \mathsf{G} \, (x-\eta) /\hbar
      \right] \, ,                    
\ee
with
\be
\rm{Im} \mathsf G  = 
      -\left(\mathsf{ U U^{\top}}  + 
             \mathsf{ V V^{\top}}
       \right)^{-1}  
       \left(\mathsf{ U U^{\top}} - 
             \mathsf{ V V^{\top}}
       \right)
             \mathsf J   \, . 
\label{ImG}
\ee    
The first observation is that when $\mathsf U = \mathsf V$ 
the quadratic term vanishes, and the phase becomes purely 
linear. 
In this well known case, the interference pattern is composed 
of straight lines parallel to the vector $\zeta$ 
joining the centers $u$ and $v$ (see Fig.~\ref{fig4}).
%
%%%%%%%%%%%%%%%%%%%%%%%%%%%%%%%%%%%%%%%%%%%%%%%%%%%%%%%%%%%%%
%%%%%%%%%%%%%%%%%%%%%%%%%%%%%%%%%%%%%%%%%%%%%%%%%%%%%%%%%%%%%
%
%%%%%%%%%%%%%%%%%%%%%%%%%%%%%%%%%%%%%%%%%%%%%%%%%%%%%%%%%%%%%
%%%%%%%%%%%%%%%%%%%%%%%%%%%%%%%%%%%%%%%%%%%%%%%%%%%%%%%%%%%%%
%

Now we switch to the general case 
$\mathsf U \ne \mathsf V$.
We want to characterize the quadratic form
$x \cdot {\rm Im} \mathsf G \, x$, 
which determines the shape of 
the oscillations of the Wigner function.
This shape will be clearly identifiable when,
by a suitable canonical coordinate transformation,
we reduce $\rm{Im} \mathsf G$ to a normal form.
%Such a normal form is the same for all superpositions 
%(\ref{Psi}).

Without loss of generality, we shall assume
$\mathsf V = \mathsf I_{2n}$. 
The general case can be recovered by setting
$\mathsf U \to \mathsf V^{-1} \mathsf U$, then making
the canonical transformation 
$x \to \mathsf V x$, etc.

Our starting point is the Euler decomposition \cite{arvind}: 
any symplectic matrix may be put in a specific diagonal 
form by the use of two orthogonal symplectic 
transformations:
\begin{equation}
\mathsf S \in {\rm Sp}(2n,\mathbb R) 
\Rightarrow
\mathsf S = \mathsf O
            \mathrm \Lambda_\mathsf{S}
            \mathsf O'                      \, ,
\end{equation}
with
$\mathsf{O,O^\prime}  \in 
{\rm Sp}(2n,\mathbb R) \cap {\rm SO}(2n)$
and
\be
\Lambda_\mathsf{S} =
{\rm diag} 
\left( s_1,\ldots,s_n,s_1^{-1},\ldots,s_n^{-1} \right),
 \,\,\, s_i \ge 1 \,.
\ee
We obtain as a corollary that $\mathsf{SS}^{\top}$ 
is diagonalized by a orthogonal symplectic matrix: 
\be
\mathsf S \mathsf S^{\top} =
            \mathsf O
            \mathrm \Lambda_{\mathsf{SS}^{\top}} 
            \mathsf O^{\top} \,,
\label{SST}
\ee
with the matrix of eigenvalues
\be
\Lambda_{\mathsf{SS}^{\top}} =
{\rm diag}
 \left(  \lambda_1,      \ldots \lambda_n,
         \lambda_1^{-1}, \ldots \lambda_n^{-1}  
\right)  \, ,
\ee
where $\lambda_i = s_i^2 \ge 1$.
We are now in the position of calculating the eigenvalues
of $\rm{Im} \mathsf G$. 
With the choice $\mathsf V = \mathsf I_{2n}$, 
Eq.~(\ref{ImG}) reduces to
\be
\rm{Im} \mathsf G^\prime  = -
  \left(
  \mathsf{ U U^{\top}} + \mathsf{ I}_{2n}
  \right)^{-1} 
  \left( 
  \mathsf{ U U^{\top}} - \mathsf{ I_{2n}} \right) 
  \mathsf J \, .                
\ee
Using the corollary (\ref{SST}) as applied to
$\mathsf{ U U^{\top}}$ 
we get
\be
-\mathsf O^{\top} \rm{Im} \mathsf G' \, \mathsf O  =
 \frac{ \Lambda_{\mathsf{ U U^{\top}}} - \mathsf{I}_{2n} } 
      { \Lambda_{\mathsf{ U U^{\top}}} + \mathsf{I}_{2n} } 
\, \mathsf J 
=
\left(
\begin{array}{cc}
 \Xi & 0_n  \\
 0_n & -\Xi
\end{array}
\right) \mathsf{J}
=
\left(
\begin{array}{cc}
 0_n & \Xi  \\
 \Xi & 0_n
\end{array}
\right)  \, ,
\ee
where $\Xi$
is a diagonal matrix with {\em positive} entries:
\be
\Xi_{ii} = 
\frac{\lambda_i - 1}{\lambda_i + 1} \equiv \theta_i \,. 
\ee 
The anti-diagonal matrix
$\mathsf O^{\top} \rm{Im} \mathsf G' \mathsf O$
can be diagonalized by the symplectic orthogonal 
matrix 
\be
\mathsf{H} = \frac{1}{\sqrt{2}}
\left(
\begin{array}{cc}
 \mathsf I_n & \mathsf I_n  \\
-\mathsf I_n & \mathsf I_n
\end{array}
\right) \, 
\ee
(a $\pi/4$ rotation in each coordinate plane), that is,
\be
   -\mathsf{H} 
    \mathsf O^{\top}
\rm{Im} \mathsf G' 
    \mathsf O 
    \mathsf{H}^{\top} 
=
\left(
\begin{array}{cc}
 \Xi & 0_n  \\
 0_n & - \Xi
\end{array}
\right)           \, .                                  
\end{equation}
So, we have shown that for any superposition (\ref{Psi})
there exists a canonical 
coordinate system $(Q,P)$ where the quadratic part of 
the phase of oscillatory term (\ref{Iosc}) reduces to 
the normal form
\be  
\label{normalform} 
\frac{1}{\hbar} 
\sum_{i=1}^n \theta_i \left( Q_i^2-P_i^2 \right) \,, 
\ee 
with $\theta_i>0$. 
In one degree of freedom, except for the degenerate
case $\mathsf U = \mathsf V$, the pattern is always
hyperbolic (see Fig.~\ref{fig4}).

The nonlinear interference patterns we described
for continuous-variable systems may also be observed 
in discrete Wigner functions of finite-dimensional
states.
Consider, for instance, the phase-space representation
of Grover's search algorithm for an $N$-qubit system 
\cite{bianucci02}.
The computer starts in a pure momentum state 
(an equal superposition of all basis states) and, 
after some iterations, evolves into a position eigenstate 
corresponding to the searched item.
At intermediate times the computer state is a weighted 
superposition of a momentum state and a position state.
The Wigner function of such states 
(see Fig.~2 in Ref.~\cite{bianucci02})
are very similar to the hyperbolic squeezed-state
superposition in our Fig.~\ref{fig4}
\cite{saraceno}.

%%%%%%%%%%%%%%%%%%%%%%%%%%%%%%%%%%%%%%%
\section{Decohered Gaussian Cat States}
\label{sec5}
%%%%%%%%%%%%%%%%%%%%%%%%%%%%%%%%%%%%%%%

This section discusses what happens to the Wigner function
of a pure Gaussian cat state when it evolves under a general 
linear dynamics, i.e., a dynamics that preserves Gaussian
states \cite{braunstein05}.
In the unitary case such dynamics are generated by
quadratic Hamiltonians.
We have seen that this corresponds to a metaplectic evolution
operator which transforms the Wigner function according to
the covariance rule (\ref{MWM})  
(if the Hamiltonian contains a term linear in $\hat x$ there 
will be some additional translation of the Wigner function).

The most general linear evolution of a density operator
$\hat \rho$ is described by the master equation
(written in the Lindblad form) \cite{dodonov}:
\begin{equation}
\frac{\partial \hat \rho }{\partial t}  = 
 \frac{1}{ i \hbar }
            \left[\hat H, \hat \rho \right] 
-\frac{1}{ 2 \hbar } 
      \sum_{k = 1 }^M \left(
        \hat L_k^\dagger \hat L_k         \hat \rho +
        \hat\rho         \hat L_k^\dagger \hat L_k -
      2 \hat L_k         \hat\rho         \hat L_k^\dagger 
                      \right)  \, ,         
\label{Lindblad}
\end{equation}
with a quadratic Hamiltonian $\hat H$ and linear Lindblad
operators $\hat L_k$,
\bea
\hat H   & = & \half \hat x^\top \mathsf B \, \hat x \, , \\
\hat L_k & = & \lambda_k \wedge \hat x            \, .
\eea
Here $\mathsf B$ is a real symmetric matrix and $\lambda_k$ 
a complex vector 
(for simplicity we assume that both
$\hat H$ and $\hat L_k$ 
are time independent).

The corresponding evolution equation for the Wigner 
function $W(x,t)$ is obtained by using the standard recipes 
for transforming master equations into partial differential 
equations \cite{barnett,oconnell03}.
In the case of a quadratic (in $\hat x$) Lindblad equation 
we arrive at a so-called linear Fokker-Plank equation:
\be
 \frac{\partial W}{\partial t}= 
-\frac{\partial}{\partial x^\top} \left( \mathsf A \, x \, W \right)
+\frac{1}{2} \frac{\partial}{\partial x^\top} \mathsf D  
 \frac{\partial}{\partial x     } W \, .
\label{fokker}
\ee
The drift and diffusion matrices, $\mathsf A$ and $\mathsf D$,
respectively, can be compactly written as 
\bea
\mathsf D & = & \hbar \, \rm{Re} \, \Upsilon \, , \\
\mathsf A & = &  
   \mathsf J \left( \mathsf B - \rm{Im} \, \Upsilon \right) \, ,
\eea
where 
\be
\Upsilon = \sum_{k=1}^M \lambda_k \lambda_k^\dagger \, .
\ee
Note that $\mathsf D$ is real symmetric nonnegative,
while, in general, $\mathsf A$ is just real 
($\rm{Im} \Upsilon$ is antisymmetric). 
In spite of some differences in the notation, 
the results above coincide with those in
Ref.~\cite{dodonov} (see also \cite{brodier04}).

Let us now consider the evolution of the Wigner function
of an initially pure Gaussian cat state.
The expression for $W(x,t)$ can be obtained 
by convolving the initial distribution $W(x_0,0)$ 
[given by Eq.~(\ref{cat})] with the 
Fokker-Planck propagator:
\be
W(x,t)= \int_0^t dx_0 \, P(x,t|x_0,0) \, W(x_0,0) \, .
\ee
The propagator $P(x,t|x_0,0)$, i.e., the Green function 
for Eq.~(\ref{fokker}), is a real Gaussian function of 
both variables $x$ and $x_0$ \cite{carmichael,dodonov}.
If $W(x_0,0)$ is a sum of Gaussians, then $W(x,t)$ will also be.
 
For an initial Wigner function like that in Eq.~(\ref{cat}), 
the integral above can be computed explicitly to get the 
desired solution at any given time $t$.
However, we are only interested in the general structure of
the Wigner function of the evolved Gaussian cat state, 
which is determined by the purely quadratic part of the 
evolving Gaussians.
For this reason we shall focus on the evolution of the 
covariance matrices of the individual Gaussians that 
compose the cat state; there are four of them: two real
ones and a complex-conjugate pair.

Given a Wigner function satisfying the Fokker-Planck equation
(\ref{fokker}), its associated covariance matrix obeys the 
following equation of motion \cite{carmichael}:
\be
\frac{d \, \mathsf C}{dt} = 
      \mathsf A \mathsf C  + 
      \mathsf C \mathsf A^\top + \mathsf D \, .
\label{covevII}
\ee
%%
%(Average values evolve independently according to
%$ d \bar{x} /dt = \mathsf A \bar{x} $.)
%%
We remark that here and below, the ``covariance" $\mathsf C$
stands for the inverse matrix of the quadratic form of any of
the four Gaussians composing the cat-state Wigner function.
At $t=0$ these matrices are symplectic but, when the dynamics
(\ref{covevII}) sets in, the symplectic symmetry is lost.

The solution satisfying the initial condition 
$\mathsf C(t=0)= \mathsf C_0$ is given by \cite{gardiner}
\be
\mathsf C(t) = e^{\mathsf A t} \,
                  \mathsf C_0  \,
               e^{\mathsf A^\top t} +
\int_0^t dt^\prime 
               e^{\mathsf A (t-t^\prime)} \,
                  \mathsf D  \,
               e^{\mathsf A^\top (t-t^\prime)}  \, .
\label{covev}
\ee
This equation allows us to evolve one-by-one 
the covariance matrices of the individual Gaussian 
terms of the cat state (\ref{cat}).

It is convenient to split Eq.~(\ref{covev}) into 
real and imaginary parts:
\bea
\label{covre}
{\rm Re} \mathsf C(t) & = & e^{\mathsf A t} \,
                      {\rm Re} \mathsf C_0  \,
                            e^{\mathsf A^\top t} +
\int_0^t dt^\prime 
               e^{\mathsf A (t-t^\prime)} \,
                  \mathsf D  \,
               e^{\mathsf A^\top (t-t^\prime)}  \, , \\
\label{covim}
{\rm Im} \mathsf C(t) & = & e^{\mathsf A t} \,
                      {\rm Im} \mathsf C_0  \,
                            e^{\mathsf A^\top t} \, . 
\eea
%
%Let us assume that the initial cat state is nondegenerate, i.e., 
If ${\rm Im} \mathsf C_0$ is nonsingular,
then, evidently,  ${\rm Im} \mathsf C(t)$ remains 
nonsingular for all times. 
Moreover, Eq.~(\ref{covim}) says that  
${\rm Im} \mathsf C(t)$ 
is {\em congruent} to
${\rm Im} \mathsf C_0$,
so, the full {\em signature} of ${\rm Im} \mathsf C(t)$ remains 
constant in time 
(this is Sylvester's law of inertia \cite{meyer};
``signature" indicates a triplet of integers:
the number of positive, negative and zero-valued eigenvalues).
Similar congruence arguments, and the fact that 
${\rm Re} \mathsf C_0$ is positive definite 
and $\mathsf D$ is nonnegative, imply that
${\rm Re} \mathsf C(t)$ remains positive for all times.

Now we have collected the basic ingredients to prove 
the main result of this section: 
the equation of motion (\ref{covev}) preserves the 
signatures of both 
${\rm Re} [\mathsf C^{-1}(t)]$ 
and 
${\rm Im} [\mathsf C^{-1}(t)]$, i.e.,
\bea
\label{teo1}
{\rm Re} [\mathsf C^{-1}(t)] 
        & \sim & {\rm Re} \left( \mathsf C_0^{-1} \right) \, \\
\label{teo2}
{\rm Im} [\mathsf C^{-1}(t)] 
        & \sim & {\rm Im} \left( \mathsf C_0^{-1} \right) \, ,
\eea
the symbol $\sim$ meaning ``congruent to".

This result leads to the following implications for 
the structure of the Wigner function that
evolves from (\ref{cat}).
Obviously, the terms that evolve from 
$W_{| \mathsf U, u \rangle}(x)$ 
and 
$W_{| \mathsf V, v \rangle}(x)$,
remain positive Gaussians for all times, 
their covariances evolving according to Eq.~(\ref{covre})
[in these cases one has ${\rm Im} \mathsf C(t)=0$, 
$\forall \, t\,$].
Concerning the interference term ${\cal I}(x)$, the preservation
of the signature of ${\rm Im} \mathsf C^{-1}(t)$ means 
that the oscillation pattern will still be represented by 
the normal form (\ref{normalform}) in some coordinate system. 
However, differently from the pure case, the new coordinates 
will not be canonical in general.
Of course, the oscillation pattern is multiplied by
a Gaussian window.

The proof of (\ref{teo1},\ref{teo2}) for the interference
term starts by expressing the 
inverse of $\mathsf C(t)$ 
in terms of its imaginary and real parts
(omitting the time dependence): 
\bea
\label{ReC}
       {\rm Re} \left( \mathsf C^{-1} \right) & = &
\left[ {\rm Re} \mathsf C +
       {\rm Im} \mathsf C        \,
       \left({\rm Re} \mathsf C \right)^{-1}   \,
       {\rm Im} \mathsf C
\right]^{-1 }       \,, \\
\label{ImC}
       {\rm Im} \left(\mathsf C^{-1}\right) & = &
\left[ {\rm Im} \mathsf C +
       {\rm Re} \mathsf C        \,
       \left({\rm Im} \mathsf C \right)^{-1}   \,
       {\rm Re} \mathsf C
\right]^{-1 }       \, .
\eea
Let us analyze the first line (\ref{ReC}): 
${\rm Re} \mathsf C>0$, 
then 
$({\rm Re} \mathsf C)^{-1}>0$, 
then
$            {\rm Im} \mathsf C      
       \left({\rm Re} \mathsf C \right)^{-1}  
             {\rm Im} \mathsf C >0 
$
(Sylvester's law). 
As the sum of positive matrices is positive, then 
$ 
{\rm Re} \left( \mathsf C^{-1} \right) 
$
is positive for all times.

In the second line (\ref{ImC}) we use that,
${\rm Re} \mathsf C$ being positive, it has a
positive square root:
${\rm Re} \mathsf C= \mathsf M^2$.
Then, defining 
$
{\rm Im} \mathsf C^\prime =
\mathsf M^{-1} ({\rm Im} \mathsf C) \mathsf M^{-1}
$,
which has the same signature as ${\rm Im} \mathsf C$,
we get
\be
{\rm Im} \left(\mathsf C^{-1}\right)  = 
\mathsf M^{-1}
 \left[ 
        {\rm Im} \mathsf C^\prime +
  \left({\rm Im} \mathsf C^\prime \right)^{-1}   \,
\right]^{-1 }  
\mathsf M^{-1}     \, .
\ee
Noting that 
$
        {\rm Im} \mathsf C^\prime +
  \left({\rm Im} \mathsf C^\prime \right)^{-1} \sim
        {\rm Im} \mathsf C^\prime
$
, we obtain
\be
{\rm Im} \left(\mathsf C^{-1}\right) \sim
{\rm Im} \mathsf C^\prime            \sim
{\rm Im} \mathsf C                   \sim
{\rm Im} \mathsf C_0                 \sim
{\rm Im} \left( \mathsf C_0^{-1} \right) \, .
\ee
The last congruence is a consequence of the initial 
correlation matrix $\mathsf C_0$
being symmetric symplectic:
$
\mathsf C_0^{-1} = \mathsf J \mathsf C_0 \mathsf J^\top
$. $\square$

%%%%%%%%%%%%%%%%%%%%%%%%%%%%%%%%%%%%%%%%%%%%%%%%%
\section{Superposition of mixed Gaussian states}
\label{sec6}
%%%%%%%%%%%%%%%%%%%%%%%%%%%%%%%%%%%%%%%%%%%%%%%%%

In the previous section we considered a pure cat state
which was subjected to the decohering action of a linear 
reservoir.
Now we study the opposite situation:  
a pure Gaussian state is first decohered and afterwards 
used as input for a cat-generating protocol.
In what respects the Wigner-function interference pattern, 
we shall see that the latter case is more general, 
as elliptical structures may also appear. 

\begin{figure}[htp]
\hspace{0.0cm}
\includegraphics[angle=0.0, width=8cm]{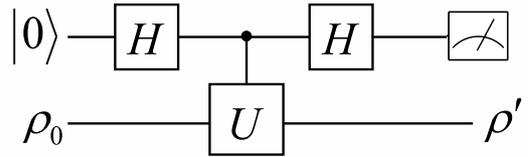}
\caption{%
Circuit for generating Gaussian superpositions.
The top line carries a single qubit initialized in the state 
$|0\rangle$.
The bottom line is fed with a continuous variable
Gaussian state $\rho_0$.
The circuit is composed of two Hadamard gates ($H$)
and a controlled-$U$ linear operation.
Measurement of the outgoing qubit produces a 
Gaussian superposition $\rho^\prime$ at the bottom output.}
\label{fig5}
\end{figure}

Consider, for instance, the cat-generating scheme of
Fig.~\ref{fig5} \cite{haroche,foot10}.
%
%%%%%%%%%%%%%%%%%%%%%%%%%%%%%%%%%%%%%%%%%%%%%%%%%%%%%%%%%%%%%
%%%%%%%%%%%%%%%%%%%%%%%%%%%%%%%%%%%%%%%%%%%%%%%%%%%%%%%%%%%%%
%
%%%%%%%%%%%%%%%%%%%%%%%%%%%%%%%%%%%%%%%%%%%%%%%%%%%%%%%%%%%%%
%%%%%%%%%%%%%%%%%%%%%%%%%%%%%%%%%%%%%%%%%%%%%%%%%%%%%%%%%%%%%
%

If the initial Gaussian state $\hat \rho$ is pure, i.e.,
$\hat \rho=|\psi \rangle \langle \psi|$, 
then the final state is also pure, 
$\hat \rho^\prime=|\psi^\prime \rangle \langle \psi^\prime|$,
with
\be
|\psi^\prime \rangle = 
              {\cal N}_\pm (1 \pm \hat U) 
              |\psi \rangle \, ,
\label{purecat}
\ee
where the sign $\pm$ depends on the result of the
qubit measurement, and ${\cal N}_\pm$ is a state-dependent
normalization
factor.
When $\hat U$ is a linear operation (a combination of a translation
and a metaplectic unitary), the output state (\ref{purecat}) 
becomes a pure Gaussian cat state, exactly like those described 
in Sec.~\ref{sec4}.
However, if the initial state is Gaussian mixed, then the
final state,
\be
\hat \rho^\prime = 
   {\cal N}_\pm (1 \pm \hat U) \, 
                       \hat \rho_0 \, 
                (1 \pm \hat U^\dagger) \, ,
\label{mixedcat}
\ee
may be called a mixed Gaussian cat \cite{jeong}.
A possible implementation of the circuit above 
for generating mixed superpositions uses two
free-propagating quantum optical modes interacting 
via a cross-Kerr nonlinear crystal \cite{jeong}.

\begin{figure}[htp]
\hspace{0.0cm}
\includegraphics[angle=-90.0, width=7cm]{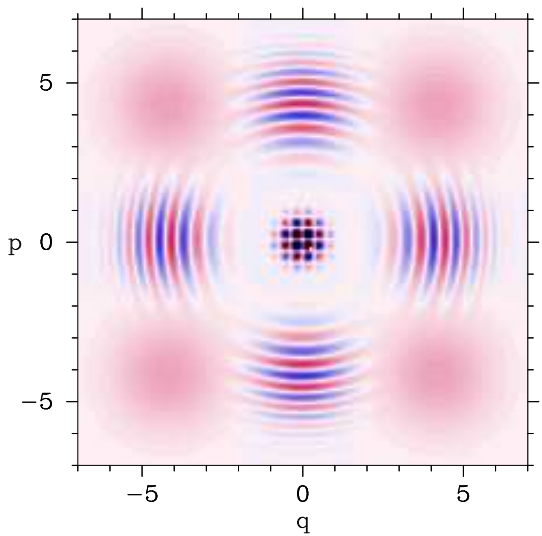}
\caption{(color online)
Wigner function of a mixed compass state.
An initial displaced thermal state is evolved with
the Kerr Hamiltonian up to a fractional revival time
($\nu=4$).}
\label{fig6}
\end{figure}

A formally similar result is produced when a single
mode propagates in a medium exhibiting a Kerr nonlinearity.
In the interaction representation, the evolution is
governed by the Hamiltonian \cite{milburn1995}
\be
\hat{H}= \gamma \hat{n}^2 \, ,
\ee
where 
$\hat{n}=\hat{a}^{\dagger}\hat{a}$ 
is the number operator, and $\gamma$ an energy scale.
At the revival time
\be
T = \frac{2\pi\hbar}{\gamma}
\ee
an initial coherent state $|\alpha_0 \rangle$ is perfectly 
reconstructed.
For times equal to $(\mu/\nu)T$, with $\mu/\nu$ an irreducible 
fraction, 
the evolved state consists of a superposition of $\nu$ 
coherent states lying on a circle of radius $|\alpha_0|$ 
(fractional revivals) \cite{revival-kerr-medium}.
However, this phenomenon is not restricted to coherent states,
because at fractional-revival times the Kerr propagator 
becomes a sum of harmonic oscillator propagators, i.e.,
\be
e^{-2 \pi i \, \mu \, \hat n^2 /\nu} = 
\sum_{k=1}^\nu c_k \, e^{-2 \pi i \, k \, \hat n/\nu}  =
\sum_{k=1}^\nu c_k \, \hat M_k \, ,
\label{fourier}
\ee
where $c_k$ are complex Fourier coefficients having equal 
moduli \cite{robinett04}. 
The notation $\hat M_k$ emphasizes the metaplectic
nature of the harmonic oscillator propagators, 
which correspond to phase-space rotations of angles 
$2 \pi k/\nu$.
Equation~(\ref{fourier}) shows that at revival times
any state will be transformed into a superposition of
$\nu$ (rotated) replicas of itself. 
If the initial state $\rho_0$ is Gaussian mixed, say a
displaced thermal state, it will evolve into the
mixed Gaussian superposition
\be
\hat \rho^\prime = 
\sum_{k,j=1}^\nu c_k c_j^\ast \, \hat M_k \, \hat \rho_0 \, 
                               \hat M_j^\dagger \, .
\label{rhokerr}
\ee
Figure~\ref{fig6}(a) exhibits an example for $\nu=4$
(a mixed ``compass" \cite{zurek01} state).
Two kind of interference patterns can be identified.
The patterns corresponding to opposite replicas are 
linear, while those corresponding to contiguous 
replicas are elliptical (circular). 
In all cases, the extension of the interference regions
are smaller than those corresponding to the diagonal
terms 
$
\hat M_k \, \hat \rho_0 \, \hat M_k^\dagger
$.
%
%%%%%%%%%%%%%%%%%%%%%%%%%%%%%%%%%%%%%%%%%%%%%%%%%%%%%%%%%%%%%
%%%%%%%%%%%%%%%%%%%%%%%%%%%%%%%%%%%%%%%%%%%%%%%%%%%%%%%%%%%%%
%
%%%%%%%%%%%%%%%%%%%%%%%%%%%%%%%%%%%%%%%%%%%%%%%%%%%%%%%%%%%%%
%%%%%%%%%%%%%%%%%%%%%%%%%%%%%%%%%%%%%%%%%%%%%%%%%%%%%%%%%%%%%
%
Had we started with a suitable thermal squeezed state we 
should have obtained hyperbolic fringes instead of elliptical 
ones (graphics not shown).

Note the formal similarity between both cat-generating schemes
described above:
The interference terms generated with the Kerr Hamiltonian 
[Eq.~(\ref{rhokerr})] can also be obtained from the scheme 
of Fig.~\ref{fig5} [Eq.~(\ref{mixedcat})], by using alternatively 
$\pi/4$ or $\pi/2$ controlled rotations.
So, in general, both schemes produce similar mixed cat states.
Such cat states are structurally different from the decohered 
cats of Sect.~\ref{sec6}, which can only exhibit hyperbolic
fringes in their Wigner functions.
 
We shall not present a detailed analytical description of the 
Wigner functions of the mixed cat states 
(\ref{mixedcat},\ref{rhokerr}).
Explicit expressions can be worked out along the lines 
of Sect.~\ref{sec4} (see also Ref.~\cite{saito96}).
However, we would like to exhibit a simple analytical
explanation of the shrinking of interference patterns 
(see Fig.~\ref{fig6}).

Consider for simplicity the binary cat produced by the Kerr 
dynamics at half the revival time ($\nu=2$):
\be
\hat \rho^\prime = 
   \frac{1}{2} (1 + i \hat M_\pi) \, \hat \rho_0 \, 
                (1 -i \hat M_\pi^\dagger) = 
   \frac{1}{2} (1 + i \hat R_0)   \, \hat \rho_0 \, 
               (1 - i \hat R_0) \, .
\ee
Here $\hat M_\pi$ denotes the half-period harmonic evolution,
which is equivalent to the parity operation $\hat R_0$, i.e., 
the reflection through the phase-space origin.
The Wigner function is calculated from Eq.~(\ref{wofx}):
\be
2 \pi \hbar \, W^\prime(x) = 
{\rm tr} \hat \rho_0 \hat R_x + 
{\rm tr} \hat R_0 \hat \rho_0 \hat R_0 \hat R_x +
2 {\rm Re} \, i \,
{\rm tr} \hat R_0 \hat \rho_0  \hat R_x \, .
\ee
Using the cyclic property of the trace, 
and the composition formulae for reflections and 
translations (\ref{RTR},\ref{TRR},\ref{RRT}), 
we obtain the cat Wigner function,
\be
2 W^\prime(x) = 
         W_0(x)+ W_0(-x) + 
             4 {\rm Re} \, i \, \chi_0(-2x) \, ,
\ee
in terms of the Wigner and characteristic functions
of the initial state $\hat \rho_0$ \cite{ozorio04}.
If $\hat \rho_0$ is a displaced thermal state, 
$\hat \rho_0 = \hat T_\eta \hat \rho_{\rm th} \hat T_\eta^\dagger$,
then we arrive at
\be
2 W^\prime(x) = 
         W_{\rm th}(x+\eta)+ 
         W_{\rm th}(x-\eta) - 
              4 \sin(x \wedge 2\eta) \, \chi_{\rm th}(2x) \, .
\ee
Remarkably the interference pattern of the
cat Wigner function is described by the characteristic 
function of the thermal state.
When temperature grows $W_{\rm th}(x)$ becomes wider.
Then, the characteristic function must shrink,
because Wigner and characteristic functions are 
related by a Fourier transform.
So, by increasing the temperature one can make the 
interference pattern as small as desired \cite{jeong}.

%For pure Gaussian states $W(x)$ and $\chi(2x)$ have 
%equal widths 
%[see Eqs.~(\ref{W0},\ref{chi0})]. 

%%%%%%%%%%%%%%%%%%%%%%%%%%%%%
\section{Concluding remarks}
\label{sec7}
%%%%%%%%%%%%%%%%%%%%%%%%%%%%%

We have studied the Wigner functions of general 
superpositions of Gaussian states.
For the pure case, we showed that the structure of the 
interference pattern is hyperbolic in general.
This structure is robust against the action of a 
linear environment.
We also analyzed two families of mixed Gaussian cat states
which may also exhibit elliptic fringes.

Our approach was geometric, qualitative.
Anyway, the analytical tools presented here can 
in principle be used for quantitative purposes,
like describing how cat states lose coherence in 
linear environments \cite{kim92,saito96,serafini05,ozorioXXX}, 
or to determine the order of nonclassicality 
\cite{vogel02} of the generalized cats.

As far as we know, generalized cat states, i.e., 
showing nonlinear intereference patterns in their Wigner
functions, have not yet 
been created in the laboratory.
Some squeezed superpositions have already been produced
\cite{ourjoumtsev07}, others may become reality soon 
\cite{liu05, marek08}. 
However, such squeezed cats are still degenerate:
as there is no relative squeezing between the superposed 
states, the Wigner interference pattern is linear.

We would like to conclude by mentioning some 
{\em theoretical} generation of ``hyperbolic" cats
\cite{saraceno}.
A set of quantum states localized on the classical periodic 
orbits of a chaotic map, can be used as a basis in 
which the description of the eigenstates of its quantum 
version is greatly simplified. 
This set can be improved with the inclusion of short time 
propagation along the stable and unstable manifolds of the 
periodic orbits \cite{ermann08}.
These ``scar functions" \cite{vergini00}, when viewed 
through a phase space representation look very much like 
the hyperbolic cat in our Fig.~\ref{fig4}(c) 
\cite{ermann08}.

%%%%%%%%%%%%%%%%%%%%%%%%%%%
\section*{Acknowledgments}
%%%%%%%%%%%%%%%%%%%%%%%%%%%
%
We thank 
A. M. Ozorio de Almeida and 
M. Saraceno
for useful discussions.
This work was supported by the 
National Institute for Science and Technology of 
Quantum Information, 
CAPES, and 
FAPERJ
(Brazilian agencies).

%%%%%%%%%%%%%%%%%%%%%%%%%%%


\begin{thebibliography}{99}
\bibitem{citation}
``Thus, standing on top of one Gaussian peak, one could still
see the other peak through the valleys 
(and between the ridges) of the interference term, even 
though its envelope is a factor of two higher than either 
Gaussian." From Ref.~\cite{zurek01}.
\bibitem{foot10}
Similar schemes for producing pure superpositions of two 
coherent states in the context of cavity QED 
are described in Ref.~\cite{toscano06}.
Pure compass-state generation in ion traps
is analyzed in Ref.~\cite{dalvit06}.
\end{thebibliography}
\end{document}